\documentclass[preprint,review,12pt]{elsarticle}
\usepackage{graphicx}

\journal{Nuclear Instruments and Methods in Physics Research A}

\begin{document}

\begin{frontmatter}

\title{Optical analysis of spherical mirrors of telescopes: the lens-less 
Schmidt case}

\author{Paolo W. Cattaneo\corref{cor1}}
\cortext[cor1]{Corresponding author}
\ead{Paolo.Cattaneo@pv.infn.it}

\address{\emph{INFN Pavia, Via Bassi 6 - 27100, Pavia, Italy }}

\begin{abstract}
The light distribution on the focal surface of spheric mirrors designed for 
telescopes in the lens-less Schmidt configuration is calculated analytically 
using geometrical optics.\\
This analysis was motivated by considerations of the design the design of the AUGER 
fluorescence detector \cite{augerprot}. Its geometrical parameters are used in the examples.
\end{abstract}

\end{frontmatter}

\section{Introduction}

The lens-less Schimdt configuration employs a circular diaphragm at the center of curvature of 
the mirror. In this configuration the coma aberration is a second order effect and the dominant 
aberration is spherical.\\ 
We are interested in calculating the light distribution on the focal surface of the spherical mirror.
This information is relevant for an optimal design of the mirror and for use in detector simulation. 

\begin{figure*}
\includegraphics[width=\textwidth]{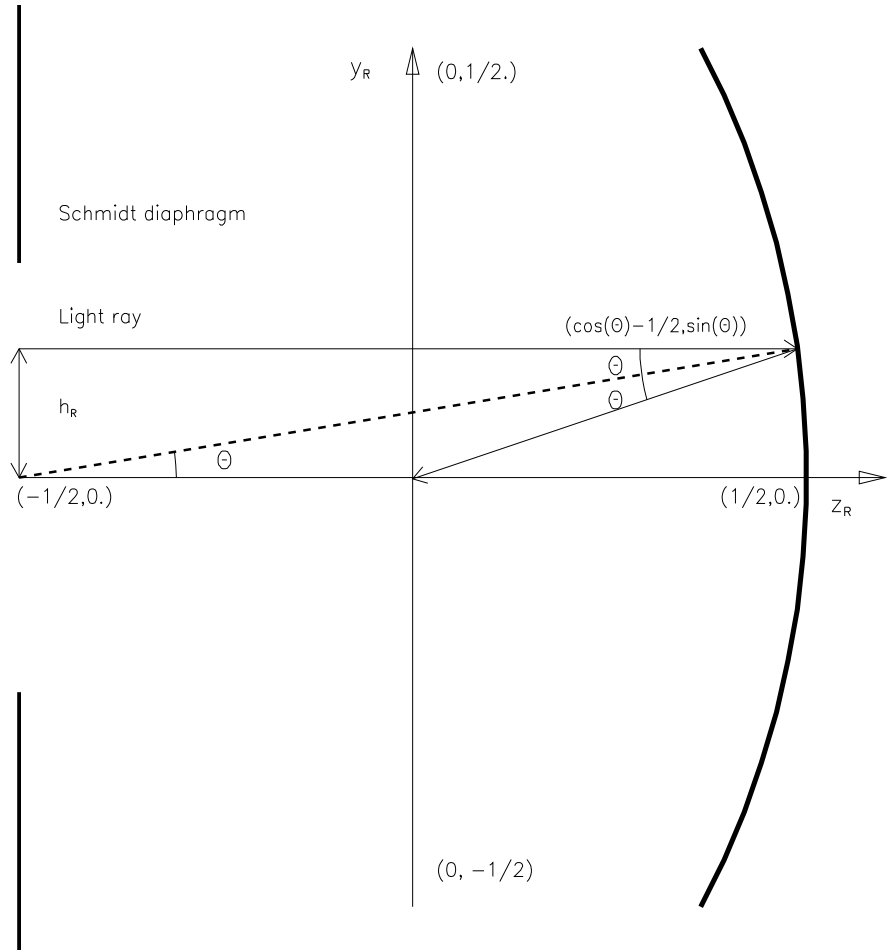}
\caption{Schematic of the mirror with the reference coordinate system.}
\label{mirror}
\end{figure*}

In Fig.\ref{mirror} a schematic of the mirror is drawn. 
The adimensional normalized coordinates $z_R=\frac{z}{R_M}$, $y_R=\frac{y}{R_M}$ and 
$\rho_R=\frac{\rho}{R_M}$ are introduced.\\
The mirror is spherical with curvature radius $R_M=3400\,\mathrm{mm}$
and size $3500\,\mathrm{mm}\times 3500\,\mathrm{mm}$ .\\
The light entrance is limited by a circular diaphragm located at the mirror center 
of curvature with a semiaperture $R_d=850\,\mathrm{mm}$ that defines 
an opening angle $\sin \Theta_M=R_d/R_M$.\\ 
The light distribution on and close to the focal surface due to paraxial rays is calculated
below including the effect of camera obscuration.

\section{Light distribution from paraxial rays without obscuration}

We consider a ray parallel to the $z_R$ axis at radial distance $h_R = h/R_M \le R_d/R_M$.
In Fig.\ref{mirror}, the ray lies in the $y_R-z_R$ plane.\\
The ray intercepts the mirror at the point $((\cos \Theta - \frac{1}{2}), \sin \Theta)$ 
where $\Theta$ is the angle with respect to the circle center (not the axis origin). 
The reflected ray forms an angle $2 \Theta$ with the $z_R$ axis. Its equation in
cylindrical coordinates $\rho_Rz_R$ is

\begin{eqnarray}
\rho_R - \sin \Theta &=& \tan2\Theta\left(z_R-\left(\cos \Theta-\frac{1}{2}
\right)\right) \nonumber\\
\rho_R &=& \tan 2\Theta \left(z_R +\frac{1}{2} \left(1 - \frac{1}{\cos \Theta}
\right)\right) 
\label{line}
\end{eqnarray}

\subsection{Calculation of the caustic}

The calculation of the light distribution on the focal surface requires the knowledge of the 
caustic of the spherical surface. The relevant formulae are in the Appendix.\\
The relevance of the caustic stems from the fact that, for $\sin \Theta \le \sin \Theta_M$, 
it is the outer envelope of the converging rays. We are interested in studying the behavior of 
the rays close to the focal surface, where the previous condition on $\Theta$ is
satisfied. Hence, the outer diameter of the light disk $\rho_R(z_R)$ is the maximum between
the radial coordinates of the caustic and of the rays from the mirror rim, 
that is Eq.\ref{line} for $\Theta = \Theta_M$.

\subsection{Light envelope and disk of minimal confusion}

The size and the intensity of the light envelope are calculated by reformulating the equation 
of the caustic and solving the following system:

\begin{eqnarray}
z_R &=& \frac{1}{2} \left[ -1 + \sqrt{1-\rho_R^\frac{2}{3}}
(1 + 2\rho_R^\frac{2}{3})\right] \nonumber \\
\rho_R &=& \tan 2\Theta_M \left(z_R +\frac{1}{2} (1 - \frac{1}{\cos \Theta_M})\right)
\end{eqnarray}

The solution is anticipated to be at $|\rho_R|<<1$, therefore the approximation in 
Eq.\ref{cauappr} in the Appendix can be applied, resulting in the following third order 
polynomial equation: 

\begin{equation}
z_R^3 - \frac{27}{64}\tan^2(2\Theta_M) \left[ z_R^2 + z_R\left(1-\frac{1}{\cos \Theta_M}
\right) + \frac{1}{4}\left(1-\frac{1}{\cos \Theta_M}\right)^2 \right] = 0.
\label{zdisk}
\end{equation}

Eq.\ref{zdisk} can be solved with the Cardano method.\\
The reflected ray from the largest angle $\Theta_M$ (mirror rim or diaphragm 
limited circle) intersects the caustic at two points. It is tangent at larger 
$z_R$, by definition of the caustic, and has a simple intersection at smaller $z_R$. 
At this point the radius of the light envelope is minimum. This is the disk of 
minimal confusion, given by

\[
p = -\frac{27}{64} \tan^2 2\Theta_M \left[ (1-\frac{1}{\cos \Theta_M}) +
\frac{9}{64} \tan^2 2\Theta_M \right]
\]
\[
q = \frac{2}{27} \left[-\frac{27}{64} \tan^2 2\Theta_M\right]^3 - \frac{1}{3}
\left[-\frac{27}{64} \tan^2 2\Theta_M\right]^2 (1-\frac{1}{\cos \Theta_M})
-\frac{27}{64} \tan^2 2\Theta_M (1-\frac{1}{\cos \Theta_M})^2
\]

\begin{equation}
z_{Rdm} = \sqrt[3]{-\frac{q}{2} + \sqrt{\frac{p^3}{27} + 
\frac{q^2}{4}}} +\sqrt[3]{-\frac{q}{2} - \sqrt{\frac{p^3}{27} + \frac{q^2}{4}}} +
\frac{9}{64}\tan^2(2\Theta_M)\nonumber\\
\label{dm}
\end{equation}

\begin{figure*}
\includegraphics[width=\textwidth]{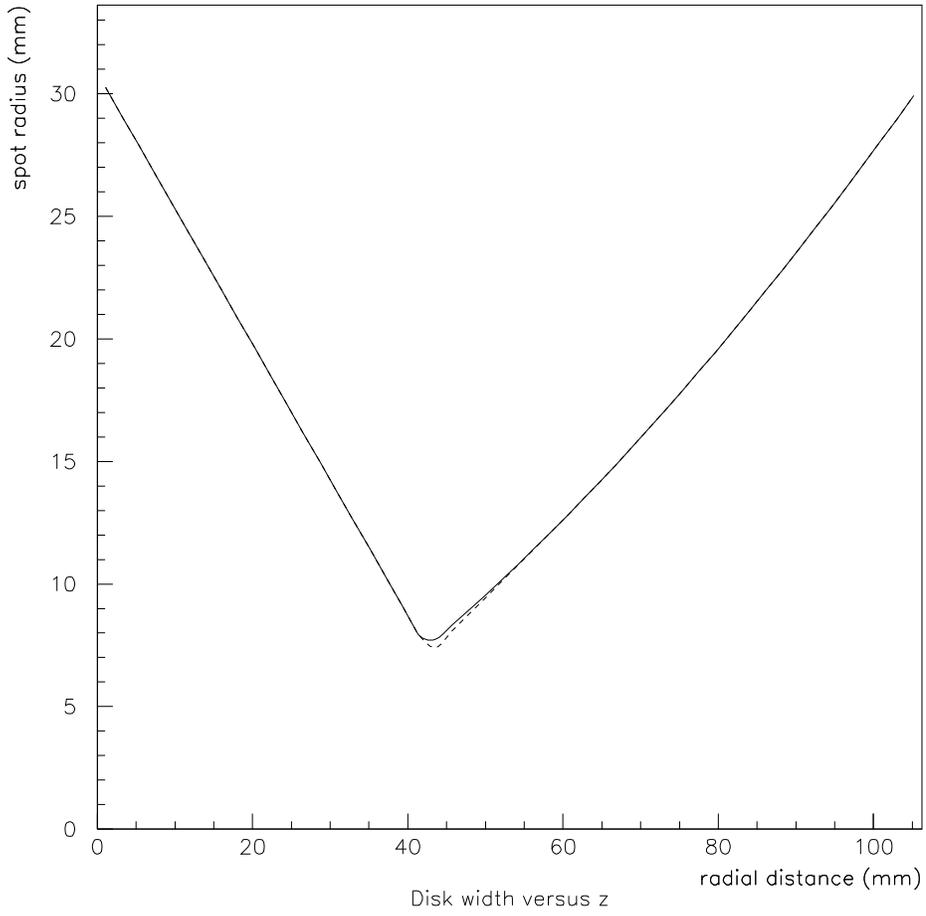}
\caption{Spot radius versus radial distance from the focus: unobscured case 
(solid), obscured case (dashed).}
\label{spot}
\end{figure*}

In Fig.\ref{focus} the envelope of light rays near the mirror focus is 
pictured together with the caustic and a bundle of rays reflected from the mirror. 
The outer envelope is given by the rays converging from the outer rim for 
\[
\frac{1}{2} (-1 + \sqrt{1-\sin^2 \Theta_M}(1+2\sin^2 \Theta_M)) \le z_R
\]
and by the caustic for 
\[
z_{Rdm}\le z_R \le \frac{1}{2} (-1 + \sqrt{1-\sin^2 \Theta_M}(1+2\sin^2 \Theta_M))
\]
and by the rays diverging from the outer rim for 
\[
z_R \le z_{Rdm}
\]
where $z_{Rdm}$ is the position of the disk of minimal confusion obtained by Eq.\ref{zdisk}.\\
The radius of the outer envelope versus the radial distance is displayed in 
Fig.\ref{spot} (solid line) for $R_M=3400\,\mathrm{mm}$. The envelope contains 
all the light reflected from the mirror.\\
The position of the disk of minimal confusion is $R=1742.52\,\mathrm{mm}$ and 
its radius $7.32\,\mathrm{mm}$.\\
In \cite{gapmp}-\cite{gapceka}, the AUGER design is analyzed using both
user-written and professional ray tracing programs. The most straightforward 
comparison with these calculations is with \cite{gapceka} where exactly the 
same geometrical parameters are used. The positions of the focal surface and the
radius of the disk of minimal confusion agree within $\approx 0.1\,\mathrm{mm}$.

\subsection{Light distribution}

\begin{figure*}
\includegraphics[width=\textwidth]{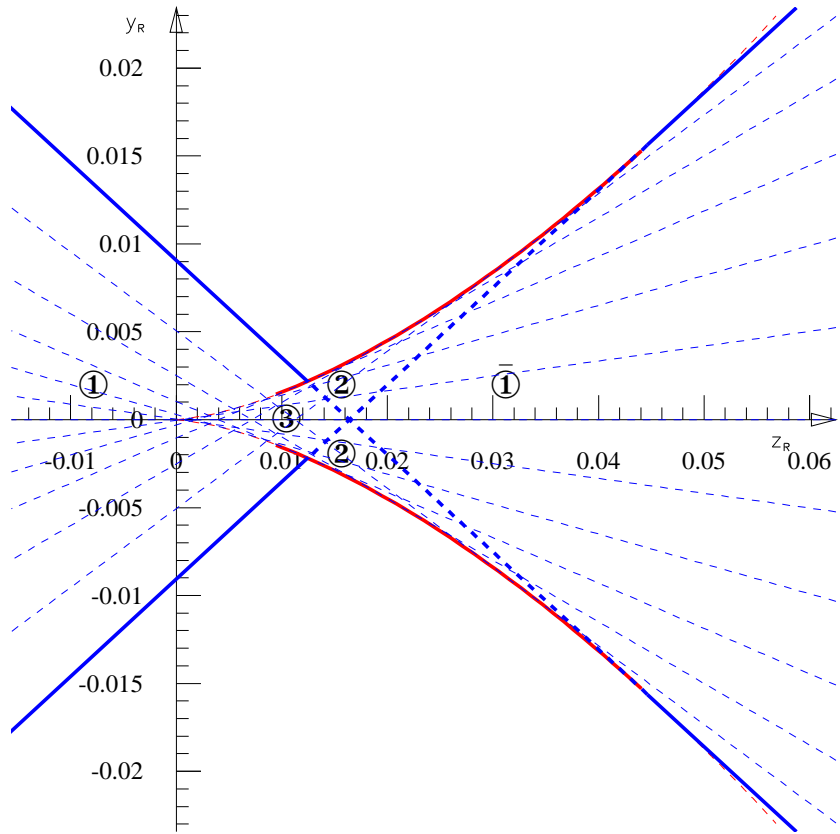}
\caption{Envelope of light rays near the focus without obscuration. Thick lines are external,
dashed internal.}
\label{focus}
\end{figure*}

The radial light distribution $\frac{df}{d\rho_R}$ at fixed $z_R$ is obtained 
by relating the known light distribution in $h_R$ to that in $\rho_R$.
The relation stems from
\begin{equation}
\frac{df}{d\rho_R} = \frac{df}{d\Theta}\frac{d\Theta}{d\rho_R} = \frac{df}{dh_R}
\frac{dh_R}{d\Theta}\frac{d\Theta}{d\rho_R} = \frac{df}{dA}\frac{dA}{dh_R}
\frac{dh_R}{d\Theta}\frac{d\Theta}{d\rho_R}
\label{dist1}
\end{equation}

The first factor is the inverse of the effective area of the mirror
$\frac{df}{dA} = (\pi R_M^2 \sin^2 \Theta_M)^{-1}$, the second is simply 
$\frac{dA}{dh_R} = 2\pi R^2_M h_R$ and the third is from $h_R=\sin \Theta$,
$\frac{dh_R}{d\Theta} = \cos \Theta$. The fourth is derived by
differentiating Eq.\ref{line} 
\[
\frac{d\Theta}{d\rho_R} = \frac{1}{\frac{2}{\cos^2 2\Theta}(z_R+\frac{1}
{2}(1-\frac{1}{\cos \Theta}))-\frac{\tan 2\Theta \sin \Theta}{2\cos^2 \Theta}}.
\]
Substituting in Eq.\ref{dist1} we find
\begin{equation}
\frac{df}{d\rho_R} = \frac{\sin 2\Theta} {
\sin^2 \Theta_M} \frac{1}{\frac{2}{\cos^2 2\Theta} (z_R+\frac{1}{2}(1-\frac{1}
{\cos \Theta})) - \frac{\tan 2\Theta \sin \Theta}{2\cos^2 \Theta}}
\label{dist2}
\end{equation}
where $\Theta$ is expressed as a function of $\rho_R$ by inverting Eq.\ref{line}.\\
If more than one $\Theta(\rho_R)$ satisfies Eq.\ref{line}, the right side of 
Eq.\ref{dist2} becomes the sum of all the solutions.\\
The inversion requires the canonical change of variable $t=\tan \Theta/2$. 
In this variable $\sin \Theta = \frac{2t}{1+t^2}$ and $\cos \Theta = \frac
{1-t^2}{1+t^2}$, so that Eq.\ref{line} becomes 
\begin{equation}
t^4 + \frac{4(z_R+1)}{\rho_R}t^3 - 6t^2 - \frac{4 z_R}{\rho_R}t + 1 =0
\label{fourth}
\end{equation}
This equation has 4 complex solutions and it is guaranteed that an even number of them are real 
although that might appear counterintuitive when Fig.\ref{focus} is analyzed graphically.\\ 
In the region (marked with $\bigcirc \!\!\!\!3$ in Fig.\ref{focus}) within the caustic and 
within the rays from 
the mirror rim after their crossing, three rays meet at a given $\rho_R$.\\ 
One is converging at small $\Theta$ and has not yet reached the caustic, one is converging 
at larger $\Theta$ and has already reached the caustic and one is diverging at even larger 
and opposite sign $\Theta$ and has already reached the caustic on the opposite site.\\
In the region (marked with $\bigcirc \!\!\!\!2$ in Fig.\ref{focus}) within the 
caustic and outside the rays from the mirror rim, there are two rays, 
the diverging ray is missing.
In the region (marked with $\overline \bigcirc \!\!\!\!1$ in Fig.\ref{focus}) within 
the caustic and within the rays from the mirror rim before their crossing, there 
is only the converging ray that has not yet reached the caustic.
In the region (marked with $\bigcirc \!\!\!\!1$ in Fig.\ref{focus}) outside the caustic and within 
the rays from the mirror rim after their crossing, there is only the diverging 
ray that has already reached the caustic.
In the region outside the caustic and outside the rays from the mirror rim, 
there are no rays.\\ 
The reason for this apparent contradiction is that Eq.\ref{fourth} for 
$\rho_R > 0$ always has one solution for $t\le -1$, that is $\Theta \le -\pi/2$, and
vice versa for $\rho_R<0$. That corresponds to a ray refracted (not reflected) by 
an unphysical hemisphere specular image around the $\rho_R$ axis of the physical 
one. In other words, Eq.\ref{line} represents full straight lines, while 
the reflected rays are only half straight lines. This
solution is unphysical and must be neglected.\\
Analysis of Eq.\ref{fourth} leads to the conclusion that, for 
$\rho_R > 0$, there is always an additional real solution for $-1<t<0$ and a 
pair of solutions that can be either both real positive or complex conjugates.\\
The reason for the presence of a region with two solutions is that the 
solution at large and opposite sign $\Theta$ lays outside the physical region 
$(-\Theta_M, \Theta_M)$. The same reason requires the region with one 
solution of converging ray at small $\Theta$.\\
In these two regions and in the region with three physical solutions, 
Eq.\ref{fourth} has four real solutions. In the region with one physical
solution outside the caustic, Eq.\ref{fourth} has two real and 
two complex solutions.\\
Eq.\ref{fourth} can be solved exactly using the canonical Ferrari approach.
Yet the solution is very cumbersome and provides little insight into 
its physical meaning.\\
The unphysical solution for the refracted ray can be removed since 
the ratio of the second to the fourth terms is $6/t^2$, where, at most, 
$t=\tan \Theta_M/2 \approx \Theta_M/2 \approx 1/8$. This ratio is at least 384 
and dropping the fourth power term will affect the physical 
solutions only by a very small amount. With this approximation,
Eq.\ref{fourth} becomes
\begin{equation}
t^3 - \frac{6\rho_R}{4(z_R+1)}t^2 - \frac{z_R}{(z_R+1)}t + \frac{\rho_R}{4(z_R+1)} =0
\label{third}
\end{equation}

\begin{figure*}
\includegraphics[width=\textwidth]{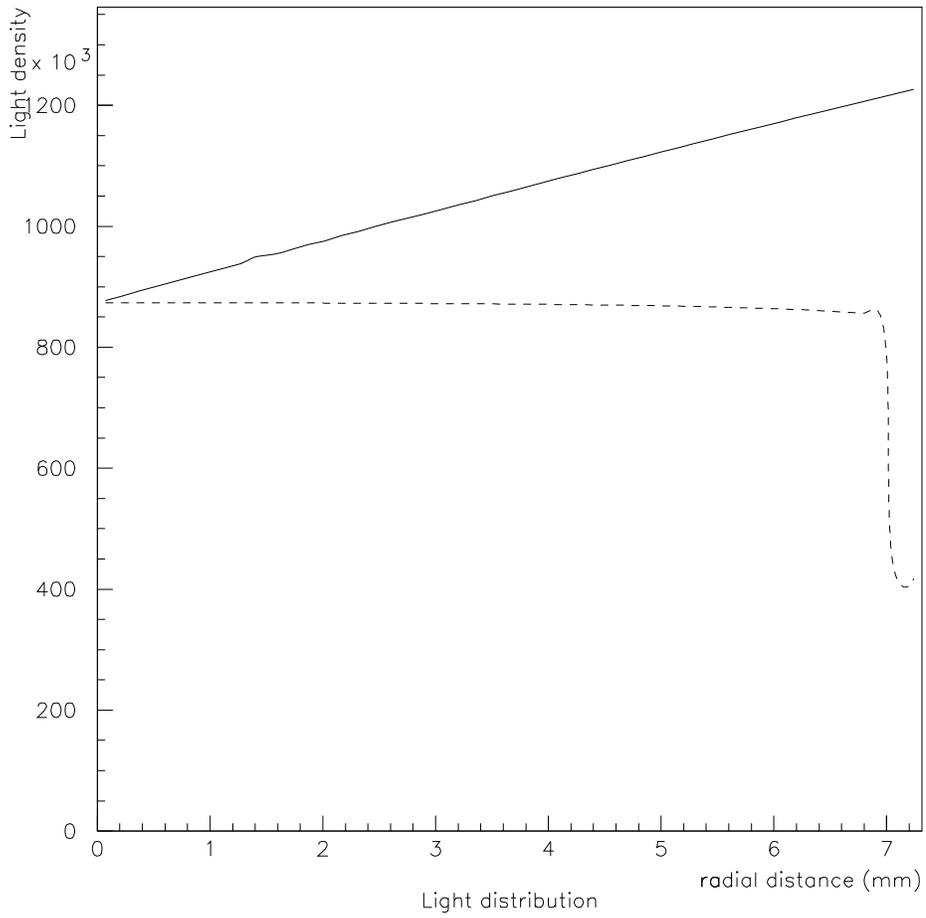}
\caption{Radial light distribution at the disk of least confusion for unobscured
(solid) and obscured (dashed) cases}
\label{lightdist}
\end{figure*}

\begin{figure*}
\includegraphics[width=\textwidth]{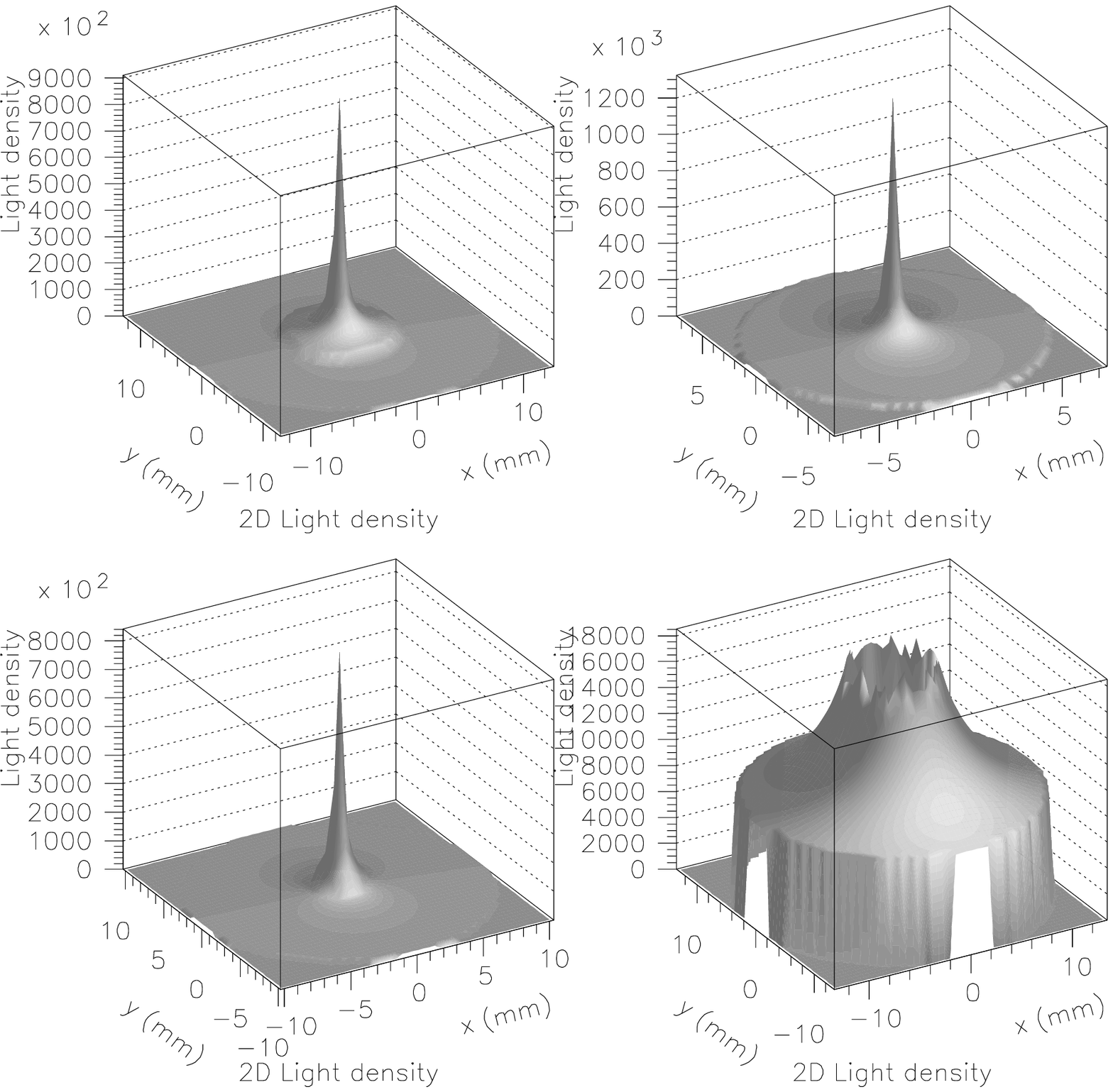}
\caption{Two dimensional light distributions at $-10,0,+10,+20\,\mathrm{mm}$ 
from the plane of least confusion}
\label{light2ddist}
\end{figure*}

The radial distribution $\frac{df}{d\rho_R}$ is plotted in Fig.\ref{lightdist}. 
The two-dimensional distribution is plotted in Fig.\ref{light2ddist}.\\
They are in good agreement with the light distributions shown in 
\cite{gapmp}-\cite{gapceka}.

\section{Light distribution from paraxial rays with obscuration}

In the AUGER design there is a $0.92\times 0.92\,\mathrm{m}^2$ photomultiplier camera 
with a spherical surface next to the mirror focal surface, 
that obscures a fraction of the incoming rays.\\
The camera does not have rotational symmetry but its
obscuration can be approximated by a disk with the same area, that is with
radius $R_c=0.519\,\mathrm{m}$ positioned next to the focal surface $z_R=1/2$. 
That implies that the rays hit the mirror only for angle $\sin \Theta \ge 
\sin \Theta_c = R_c/R_M$, that is for $\Theta \ge \Theta_c = 8.78^\circ$.

\subsection{Light envelope and disk of minimal confusion}

\begin{figure*}
\includegraphics[width=\textwidth]{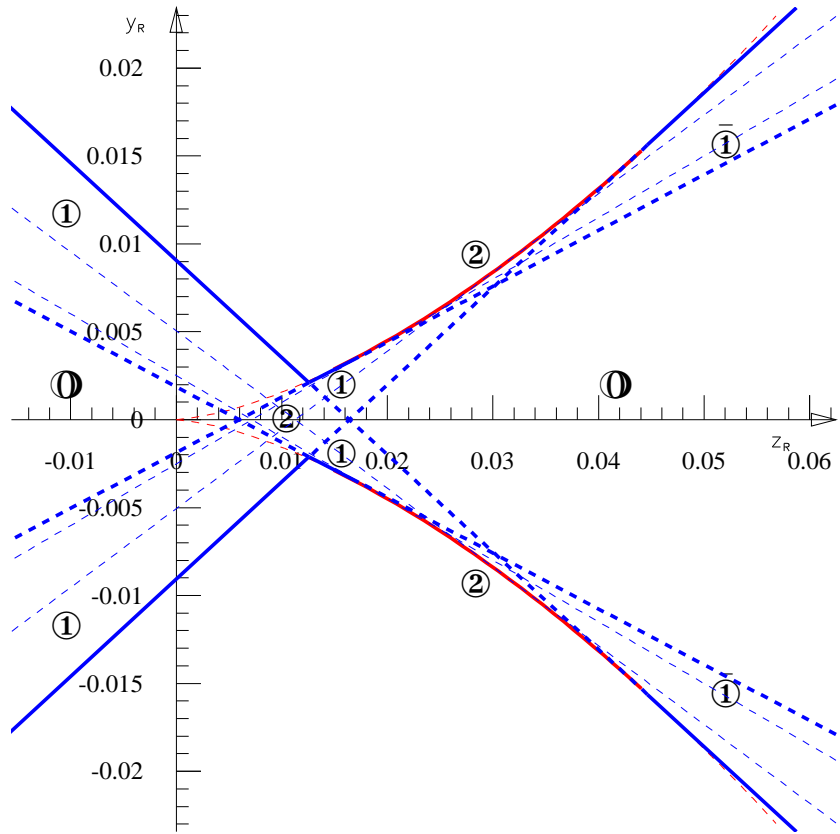}
\caption{Envelope of light rays next to the focus with obscuration. Thick lines 
are external, dashed internal.}
\label{focus_obsc}
\end{figure*}

The obscuration changes the light distribution along $z_R$ and might change the 
size and position of the disk of minimal confusion. The outer envelope of reflected 
light can be deduced by referring to Fig.\ref{focus_obsc} and noting that the 
light rays close to the $z_R$ axis are obscured. Hence the outer light envelope is 
given by the converging rays from the mirror rim up to the caustic, then by the 
caustic, then, possibly, by the converging rays from the rim of the obscuration
disk, then by the diverging rays from the mirror rim.\\
It depends on the obscuration area if the converging rays from the rim of the 
obscuration disk intersect the caustic at $z_{Rc}(\Theta_c)$ before or after 
$z_{Rdm}(\Theta_M)$ from Eq.\ref{dm}, that is if
\[
z_{Rdm}(\Theta_M) \ge \frac{1}{2}(-1+\sqrt{1-\sin^2\Theta_c}(1+2\sin^2 \Theta_c))
\approx \frac{3}{4}\sin^2\Theta_c = z_{Rc}(\Theta_c)
\]
where the caustic approximation of Eq.\ref{cauappr} is used.\\
Therefore if $\sin \Theta_c \le \sqrt{z_{Rdm}(\Theta_M) 4/3}$, the outer envelope
is unchanged, otherwise for $z_{Rdm}<z_R<z_{Rc}$ it is given by the converging rays
from the rim of the obscuration disk. In this case the disk of minimal 
confusion is obtained by the intersection of the diverging ray from the outer rim with
the converging one on the opposite side from the rim of the obscuration disk.
The result is

\begin{eqnarray}
z_{Rdm} &=& \frac{1}{2}\left(-1 + \left(\frac{\tan 2\Theta_M}{\cos \Theta_M} +
\frac{\tan 2\Theta_c}{\cos \Theta_c}\right)
\frac{1}{\tan 2\Theta_c+\tan 2\Theta_M}\right) \nonumber \\
\rho_{Rdm} &=& \frac{\tan 2\Theta_c}{2} \left(\left(\frac{\tan 2\Theta_c}
{\cos \Theta_M}+\frac{\tan 2\Theta_c}{\cos \Theta_c}\right)
\frac{1}{\tan 2\Theta_M+\tan 2\Theta_c}-\frac{1}{\cos \Theta_c}\right)
\end{eqnarray}

This condition is verified in the AUGER design and the location of the disk of 
minimal confusion changes slightly. Its position is $R=1742.80\,\mathrm{mm}$ 
and its radius is $7.17\,\mathrm{mm}$. The radius of the envelope versus the 
radial distance for the obscured case is displayed in Fig.\ref{spot} (dashed 
line). The difference between it and that of the unobscured case is minimal.\\

\subsection{Light distributions}

The results of the unobscured case are still valid, provided that only solutions
with $\Theta_c \le |\Theta| \le \Theta_M$ are considered.\\
The radial light distributions for the obscured case is presented in 
Fig.\ref{lightdist} .

\section{Conclusions}

Detailed analytical calculations of the spherical aberration of a telescope 
spherical mirror designed for detecting fluorescence light emitted in the atmoshpere 
by extended air showers generated by very high energy cosmic ray 
interactions are presented.
The position of the focal surface of the mirror and the light distribution on 
and close to it have been calculated.
The calculations are useful both for the design of the detector and for the detector 
simulation. The AUGER Fluorescence Detector has been used as example but the results 
are applicable to any similar detector.

\section*{Appendix}

\subsection*{Third order equation}

We recall the solution of the third order polynomial equation. The original 
equation and the intermediate steps leading to the solution 
$x_1$, $x_2$ and $x_3$ are listed (where $\omega = \sqrt[3]{1}$, $\omega \ne 1$)

\[
0 = x^3 + a_1x^2 + a_2x + a_3 
\]
\[
p = a_2-\frac{a_1^2}{3} \qquad q = \frac{2a^3_1}{27}-\frac{a_1a_2}{3} +a_3 
\]
\[
P = \sqrt[3]{-\frac{q}{2} + \sqrt{\frac{p^3}{27} + \frac{q^2}{4}}} 
\]
\[
Q = \sqrt[3]{-\frac{q}{2} - \sqrt{\frac{p^3}{27} + \frac{q^2}{4}}} 
\]
\begin{equation}
x_1 = P+Q - \frac{a_1}{3}\quad x_2 = \omega P+ \omega^2 Q - \frac{a_1}{3} \quad
x_3 = \omega^2 P+\omega Q - \frac{a_1}{3} 
\label{solut}
\end{equation}

Two roots coincides if and only if $\frac{p^3}{27} + \frac{q^2}{4}=0$. In this case, the
third distinct root is $x_1$, that is the solution determining the disk of least confusion 
in our application.

\subsection*{Some formulae about the caustic}

The calculation of the analytical form of the caustic from a spherical surface is a 
standard calculation, but it is surprisingly difficult to find it in text books. 
Following \cite{rossi}, the caustic is the locus of intersection of neighbouring reflected 
rays when their distance becomes infinitesimal. Referring to the coordinate 
system in Eq.\ref{mirror} and using the dimensionless variables $z_R = \frac{z}{R_M}$ and 
$y_R=\frac{y}{R_M}$, it can be calculated by equating the derivative of Eq.\ref{line} 
with respect to $\Theta$ to 0 and solving the system

\begin{eqnarray}
y_R &=& \tan 2\Theta \left(z_R +\frac{1}{2} \left(1 - \frac{1}{\cos \Theta}\right)
\right) \nonumber\\
0 &=& \frac{2}{\cos^2 2\Theta} \left(z_R+\frac{1}{2}\left(1-\frac{1}{\cos\Theta}
\right)\right)
-\frac{1}{2}\frac{\tan 2\Theta \tan \Theta}{\cos \Theta}
\end{eqnarray}

The resulting curve can be expressed in parametric form as 

\begin{eqnarray}
y_R(\Theta) &=& \sin^3 \Theta \nonumber\\
z_R(\Theta) &=& \frac{1}{2}(-1+\sqrt{1-\sin^2\Theta}(1+2\sin^2 \Theta))
\label{caustic}
\end{eqnarray}

The caustic in three dimensions is 
obtained by rotating the curve in Eq.\ref{caustic} around the $z_R$ axis.\\
It is useful to express exactly the caustic in alternative forms as 
$z_R(y_R)$, $y_R(z_R)$ and relating the parameter $\Theta$ to $y_R$ and $z_R$.
From Eq.\ref{caustic}, 

\begin{eqnarray}
\Theta(y_R) &=& \arcsin \left(\sqrt[3]{y_R}\right) \nonumber\\
z_R(y_R) &=& \frac{1}{2}(-1+\sqrt{1-y_R^{2/3}}(1+2y_R^{2/3})).
\end{eqnarray}

A common and useful approximation for $|\sin \Theta|\ll 1$ (equivalent to
$|y_R|\ll 1$ and $|z_R|\ll 1$) is obtained from $\sqrt{1-y_R^\frac{2}{3}} \approx
1 -\frac{1}{2} y_R^\frac{2}{3}$, that implies
\begin{eqnarray}
z_R(y_R) &=& \frac{3}{4} y_R^\frac{2}{3} \nonumber \\
y_R(z_R) &=& \pm (\frac{4z_R}{3})^\frac{3}{2} \nonumber \\
\Theta(z_R) &=& \pm \arcsin \left(\sqrt{\frac{4z_R}{3}}\right) 
\label{cauappr}
\end{eqnarray}

The exact expression for $y_R(z_R)$ or, equivalently, for $\Theta(z_R)$ is obtained 
from Eq.\ref{caustic} by writing
\[
z_R(\Theta) = \frac{1}{2}(-1+\cos\Theta(3-2\cos^2 \Theta))
\]
which can be reformulated as a third order polynomial equation in $\cos\Theta$
\[
\cos^3 \Theta - \frac{3}{2}\cos\Theta + (z_R + \frac{1}{2}) = 0.
\]
The discriminant of this equation is 
\[ 
\Delta = -\frac{1}{8}+\frac{1}{4}(z_R+\frac{1}{2})^2
\]
which is negative for $-\frac{\sqrt{2}+1}{2}<z_R<\frac{\sqrt{2}-1}{2}=z_{R0}$. 
For these values, there are three real solutions:

\begin{eqnarray}
\cos \Theta_1(z_R) &=& \sqrt{2} \cos\left 
(\frac{3\pi+\arctan\left(\frac{\sqrt{2-4(z_R+\frac{1}{2})^2}}{2(z_R+\frac{1}{2})}\right)}{3} \right) \nonumber \\
\cos \Theta_2(z_R) &=& \sqrt{2} \cos\left( 
\frac{5\pi+\arctan\left(\frac{\sqrt{2-4(z_R+\frac{1}{2})^2}}{2(z_R+\frac{1}{2})}\right)}{3} \right) \nonumber \\
\cos \Theta_3(z_R) &=& \sqrt{2} \cos\left( 
\frac{\pi+\arctan\left(\frac{\sqrt{2-4(z_R+\frac{1}{2})^2}}{2(z_R+\frac{1}{2})}\right)}{3} \right) 
\end{eqnarray}

The physical meaning of these solutions can be understood by considering them for $z_R=0$

\begin{eqnarray}
\cos \Theta_1(0) &=& \sqrt{2} \cos (\frac{11}{12}\pi) = -\frac{\sqrt{2}}{2} 
\sqrt{2+\sqrt{3}} \nonumber \\ 
\cos \Theta_2(0) &=& \sqrt{2} \cos (\frac{\pi}{4}) = 1 \nonumber \\ 
\cos \Theta_3(0) &=& \sqrt{2} \cos (\frac{5}{12}\pi) = \frac{\sqrt{2}}{2} 
\sqrt{2-\sqrt{3}}
\end{eqnarray}
The first is unphysical because the modulus is larger than one, the second 
corresponds to the tip of the caustic and the third is a large angle solution.\\
Depending on the quadrant, the solutions are always monotonic versus 
$z_R$ either increasing ($\Theta_1$ and $\Theta_3$) or decreasing ($\Theta_2$) 
within the range $-\frac{1}{2} \le z_R \le z_{R0}$. The lower bound of the range is
physical, the upper one comes from requiring three real solutions. For $z_R=z_{R0}$,
$\Theta_2 = \Theta_3 = \pi/4$ and for $z_R>z_{R0}$, $\Theta_2$ and $\Theta_3$ become 
complex conjugate.\\
The solution relevant for the application to a mirror is therefore $\Theta_2$
and we can write the exact parametrization of the caustic for $0\le z_R \le z_{R0}$

\begin{eqnarray}
\Theta(z_R) &=& \pm \arccos \left(\sqrt{2} \cos\left (\frac{5\pi+\arctan(\frac
{\sqrt{2-4(z_R+\frac{1}{2})^2}}{2(z_R+\frac{1}{2})})}{3}\right)\right)\nonumber \\
y_R(z_R) &=& \pm\left(-\cos2\left(\frac{5\pi+\arctan(\frac{\sqrt{2-4(z_R+\frac{1}{2})^2}}
{2(z_R+\frac{1}{2})})}{3} \right)\right)^\frac{3}{2} \nonumber 
\end{eqnarray}

The large angle solution is $\Theta_3$ and reads

\begin{eqnarray}
\Theta(z_R) &=& \pm \arccos \left(\sqrt{2} \cos\left (\frac{\pi+\arctan(\frac
{\sqrt{2-4(z_R+\frac{1}{2})^2}}{2(z_R+\frac{1}{2})})}{3}\right)\right)\nonumber \\
y_R(z_R) &=& \pm\left(-\cos2\left(\frac{\pi+\arctan(\frac{\sqrt{2-4(z_R+\frac{1}{2})^2}}
{2(z_R+\frac{1}{2})})}{3} \right)\right)^\frac{3}{2} \nonumber 
\end{eqnarray}

\section*{Acknowledgments}

We warmly acknowledge Prof. William Molzon from the Department of Physics and Astronomy of the University
of California, Irvine, CA, USA, for his careful proofreading of the manuscript.

\bibliographystyle{unsrt}
\bibliography{caustic}

\end{document}